\documentclass{optica-article}

\journal{opticajournal} 

\articletype{Research Article}

\usepackage{lineno}

\usepackage[super]{nth}
\usepackage{graphicx}
\usepackage{dcolumn}
\usepackage{bm}
\usepackage{upgreek}
\usepackage{xcolor}
\usepackage{enumitem}
\usepackage{placeins}
\usepackage{braket}
\usepackage{siunitx}

\usepackage{dsfont}
\usepackage{comment}

\begin{document}

\title{Frequency shifting of light via multiple ultrasound waves in scattering media} 

\author{Adam Kinos}
\address{Department of Physics, Lund University, P.O. Box 118, SE-22100 Lund, Sweden}
\address{Deep Light Vision AB, Gasverksgatan 3A, SE-22929 Lund, Sweden}
\email{adam.kinos@fysik.lth.se}

\begin{abstract*}
We derive analytical expressions to describe how light is frequency-shifted when interacting with ultrasound within scattering media, due to the modulation of the refractive index induced by the ultrasound pressure waves. The model is validated through Monte Carlo simulations, works for high ultrasound pressures, and allows for many simultaneous ultrasound waves or frequency components, which is important due to the non-linear propagation effects in tissue. We also provide critical insights into how the ultrasound properties can be optimized for an enhanced efficiency of the light to be frequency-shifted, facilitating applications in ultrasound optical tomography and other photonic diagnostic techniques. 
\end{abstract*}

\section{\label{sec:intro}Introduction}
Understanding how light and ultrasound interact in scattering media is important to drive the development of biomedical imaging techniques such as ultrasound optical tomography. The multiple scattering events of light lead to complex paths and diffusion-like behavior. When ultrasound is introduced, it induces periodic variations in the medium’s refractive index and mechanical displacement of scatterers, and it can even change the optical properties of the medium, although this last effect is often negligible \cite{Wang2001b}. These periodic variations can cause frequency shifts in the light, a phenomenon that ultrasound optical tomography utilizes by filtering out and examining only the frequency-shifted light, which must have originated from the ultrasound location. This leads to spatially resolved optical contrast measured non-invasively at a deeper level than conventional optical methods \cite{Wang1998, Elson2011, Gunther2017}.

Several analytical and numerical models exist to study the light-ultrasound interaction \cite{Leutz1995, Kempe1997, Mahan1998, Wang2001b, Wang2001a, Sakadzic2005, Blonigen2005, Lai2009, Laudereau2016, Bengtsson2019, Huang2020}. However, most studies have focused on low acoustic pressures and one ultrasound frequency, but for ultrasound optical tomography the pressures are high (in the range of MPa \cite{Hill2021}), and then non-linear ultrasound propagation effects are also important \cite{Duck2002}. This was recently addressed in Ref. \cite{Hill2021}, which presented a numerical model and validated it against experimental measurements. 

In this article, we complement previous works by developing a comprehensive analytical framework for describing the frequency content of light modulated by ultrasound in scattering media that works for high ultrasound pressures and multiple ultrasound frequency components, and we validate the model through Monte Carlo simulations. 

This analytical model not only enhances the understanding of light-ultrasound interactions but also offers a tool for optimizing ultrasound parameters to maximize the efficiency for light to be frequency-shifted. Such optimization could be important for improving the efficacy of ultrasound optical tomography and similar modalities, where precise control over light modulation is essential for accurate imaging and analysis.

The article is structured as follows: Section \ref{sec:theory} presents the theoretical framework for light-ultrasound interaction, while Sec. \ref{sec:results} compares the analytical predictions with Monte Carlo simulations. Finally, Sec. \ref{sec:conc} concludes with a summary of the findings and implications for future research. To ensure that this article is easy to follow, Table \ref{tab:parameters} provides a list of the most important parameters used in this work. 

\begin{table}
\caption{\label{tab:parameters}Descriptions and symbols of the most important parameters used in this work.}
\begin{tabular}{|l l|l l|}
\hline
\textbf{Description}&\textbf{Symbol}&\textbf{Description}&\textbf{Symbol}\\
\hline
Ultrasound pressure & $P(t,\boldsymbol{r})$ & Relative light intensity at $n\omega_1$ & $I_\text{rel}(n)$ \\
Ultrasound frequencies & $\omega_m$ & Order of frequency-shift & $n$ \\
Ultrasound wave vectors & $k_m$ & Bessel function of order $n$ & $\mathcal{J}_{n}(x)$ \\
Ultrasound phases & $\alpha_m$ & Random photon path & $\gamma$\\
Ultrasound pressure amplitudes & $\Gamma_m$ & Number of straight sections in path & $J$ \\
Time-dependent refractive index & $n(t,\boldsymbol{r})$ & Path integral parameters (Sec. \ref{sec:path_integral}) & $B_m$, $\beta_m$  \\
Background refractive index & $n_0$ & Light-ultrasound interaction & \\
Piezo-optic coefficient & $\frac{dn}{dP}$ & strength for path (Eq. (\ref{eq:x_m})) & $x_m$ \\
Scattering coefficient & $\mu_s$ & Distance traveled within &\\ 
Absorption coefficient & $\mu_a$ & ultrasound volume & $L$ \\
Electric field & $E(t)$ & Probability density function of $L$ & $\mathcal{P}(L)$  \\
Vacuum wave vector of the light & $\kappa_0$ & Probability of the number of straight & \\
Induced electric field phase & $\phi(t)$ &  sections within ultrasound volume & $P(J)$ \\
\hline
\end{tabular}
\end{table}

\section{Light-ultrasound interaction}\label{sec:theory}
Two main effects cause light to interact with an ultrasound focus and become frequency-shifted. The ultrasound can cause scatters to move, which alters the physical path length the photon travels, or it affects the refractive index, which changes the optical path length. In this study, we focus on the second effect and assume that the first is negligible, in accordance with the results of Ref. \cite{Hill2021}. Previous studies have also found that the movement of scatters has less importance if the ultrasound wave vector multiplied by the mean free path of the scattered photons is much greater than one \cite{Wang2001a, Gunther2017}, which is the case for the MHz ultrasound frequencies and reduced scattering coefficients of around $7.5$ cm$^{-1}$ that are typical values for examining tissue using ultrasound optical tomography. 

We assume that the ultrasound pressure can be described as several sinusoidal plane wave oscillations, with frequencies $\omega_m$, wave vectors $\boldsymbol{k}_m$, phases $\alpha_m$, and pressure amplitudes, $\Gamma_m(t, \boldsymbol{r})$;
\begin{align}\label{eq:pressure}
    P(t, \boldsymbol{r}) &= \sum_m \Gamma_m(t, \boldsymbol{r}) \sin(\omega_m t - \boldsymbol{k}_m\cdot \boldsymbol{r} - \alpha_m).
\end{align}
Similar to previous theoretical works \cite{Wang2001b, Wang2001a, Sakadzic2005, Blonigen2005, Laudereau2016, Huang2020, Hill2021}, we assume that the refractive index changes linearly with the pressure according to 
\begin{align}\label{eq:refractive_index}
    n(t, \boldsymbol{r}) &= n_0 + n_0 \frac{dn}{dP} P(t, \boldsymbol{r}), 
\end{align}
where $\frac{dn}{dP}$ is the piezo-optic coefficient. 

Given a random photon path $\gamma$, the phase due to the time-varying optical path length is given by 
\begin{align}\label{eq:phase}
   \phi(t) = \kappa_0 \int_\gamma n(t, \boldsymbol{r}) ds.
\end{align}
where $\kappa_0$ is the vacuum wave vector of the light. This phase modulates the electric field as
\begin{align}\label{eq:electric_field}
    E(t) = E_0(t) e^{i \phi(t)}, 
\end{align}
where $E_0(t)$ is the unperturbed electric field of the incoming light. 

The goal of this work is to describe the frequency content of the perturbed electric field $E(t)$, for any given ultrasound frequency, pressure, and volume, and any optical scattering coefficient of the underlying medium. However, to do this, we will assume that the pressure amplitudes are constant in the fixed volume where the light and ultrasound interact, i.e., $\Gamma_m(t, \boldsymbol{r}) = \Gamma_m$.

\subsection{Frequency content of the electric field}\label{sec:frequency_content}
To obtain the electric field frequency content obtained from Eqs. (\ref{eq:pressure})-(\ref{eq:electric_field}), we start by noting that the time-independent background refractive index term, $n_0$, in Eq. (\ref{eq:refractive_index}) will not alter the frequency content of the light, and can therefore be ignored in the subsequent analysis. The final expression then involves integrals on the following form
\begin{align}
    \int_\gamma \sin(\omega_m t - \boldsymbol{k}_m\cdot \boldsymbol{r} - \alpha_m) ds  &= \frac{1}{2i} \left(e^{i\omega_m t} \int_\gamma e^{-i(\boldsymbol{k}_m\cdot \boldsymbol{r} + \alpha_m)}ds - e^{-i\omega_m t} \int_\gamma e^{i(\boldsymbol{k}_m\cdot \boldsymbol{r} + \alpha_m)}ds \right)\nonumber \\
    &= B_m \sin(\omega_m t - \beta_m), \nonumber \\
    &B_m e^{-i\beta_m} = \int_\gamma e^{-i(\boldsymbol{k}_m\cdot \boldsymbol{r} + \alpha_m)}ds, 
\end{align}
where the solution to the path integral on the last line is discussed further in Sec. \ref{sec:path_integral}. 

Thus, the problem is reduced to obtaining the frequency content of
\begin{align}\label{eq:x_m}
    E(t) = E_0(t) \prod_m e^{i x_m \sin(\omega_m t - \beta_m)}, \quad x_m = \kappa_0 n_0 \frac{dn}{dP} \Gamma_m B_m.
\end{align}
From the convolution theorem, we know that the Fourier transform of $E(t)$ is the convolution between the Fourier transforms of the individual factors $E_0(t)$ and $\prod_m e^{i x_m \sin(\omega_m t - \beta_m)}$. Thus, we are interested in the frequency content of the second factor, which can be rewritten using a Fourier series expansion of the periodic function
\begin{align}
    \prod_m e^{i x_m \sin(\omega_m t - \beta_m)} = \prod_m \left( \sum_{n=-\infty}^{\infty} c_{m,n} e^{i n (\omega_m t - \beta_m)} \right) \nonumber \\
    c_{m,n} = \frac{1}{2\pi}\int_{-\pi}^{\pi} e^{i x_m \sin(\tau)} e^{i n \tau} d\tau = \mathcal{J}_{n}(x_m), \quad \tau = \omega_m t - \beta_m,
\end{align}
where $\mathcal{J}_{n}(x_m)$ is the Bessel function of the first kind of order $n$, and in our case $n$ also represents the order of frequency-shift, e.g., for $n=2$ the light is shifted by twice the ultrasound frequency $\omega_m$. 

While the solution is valid for ultrasounds with arbitrary frequency components, we will now focus on the case where the ultrasound consists of frequency harmonics of a fundamental frequency $\omega_1$. Thus, the expression becomes
\begin{align}
    \prod_m \left( \sum_{n=-\infty}^{\infty} \mathcal{J}_{n}(x_m) e^{i n (m \omega_1 t - \beta_m)} \right) &= \sum_{n=-\infty}^{\infty} \left[ \sum_{\{n_m\} \in \mathcal{N}(n)} \prod_m  \mathcal{J}_{n_m}(x_m) e^{-i n_m \beta_m}\right] e^{i n \omega_1 t}, \nonumber \\
    \mathcal{N}(n) &= \left\{ \{n_m\} \mid \sum_m n_m m = n \right\}, 
\end{align}
where $\mathcal{N}(n)$ is the set of all vectors $\{n_m\} = [n_1, n_2, ..., n_m]$ fulfilling the requirement $\sum_m n_m m = n$, and $n$ now specifically refers to being frequency-shifted by $n\omega_1$. For example, when interested in the amplitude of being frequency-shifted by $+\omega_1$, that is $n = 1$, and there only exists the fundamental and first overtone, that is $m$ is limited to $1$ and $2$, a few elements in $\mathcal{N}(1)$ are $\{n_m\} = [n_1, n_2] = [1, 0]$, $[3, -1]$, and $[-1, 1]$.

The relative light intensities, which represent the fraction of the intensity at different frequencies $n \omega_1$, are then given by
\begin{align}\label{eq:I_rel}
    I_\text{rel}(n) = \left| \sum_{\{n_m\} \in \mathcal{N}(n)} \prod_m  \mathcal{J}_{n_m}(x_m) e^{-i n_m \beta_m} \right|^2, 
\end{align}
where $x_m$ is given in Eq. (\ref{eq:x_m}), and $B_m$ and $\beta_m$ are calculated in the next section.

\subsection{Path integral}\label{sec:path_integral}
To obtain $B_m$ and $\beta_m$, we must solve path integrals on the form
\begin{align}
    B_m e^{-i\beta_m} &= \int_\gamma e^{-i(\boldsymbol{k}_m\cdot \boldsymbol{r} + \alpha_m)}ds. 
\end{align}
For this work, we assume that $\boldsymbol{k}_m\cdot \boldsymbol{r} = k_{m, z} z$, i.e., that all ultrasound frequency components travel in the same direction. However, the framework can be readily extended to arbitrary directions by defining a new coordinate system for each ultrasound frequency component, aligning the new $z$-axis to the propagation axis, and then transforming the path $\gamma$ into this new coordinate system.  

The path $\gamma$ the photon takes is a random scattering path through the ultrasound volume where the pressure is $\Gamma_m$, and thus depends on the ultrasound focus size and the scattering coefficient of the medium. We model the path as $J$ straight lines from the initial position $\boldsymbol{r}_0$ via $J-1$ scattering events, $\boldsymbol{r}_j$, to the final position, $\boldsymbol{r}_J$. 
\begin{align}
    B_m e^{-i\beta_m} \approx \sum_{j=1}^J \int_{\gamma_j} e^{-i (k_{m, z} z + \alpha_m)} ds = i e^{-i\alpha_m} \sum_{j=1}^J \Delta L_j \frac{e^{-i k_{m, z} \left(z_0 + \sum_{l=1}^{j-1} \Delta z_l\right)}}{k_{m, z} \Delta z_j} \left( e^{-i k_{m, z} \Delta z_j} - 1 \right), 
\end{align}
where $\Delta L_j = |\boldsymbol{r}_j - \boldsymbol{r}_{j-1}|$ is the length of the $j^{\text{th}}$ straight line, $\Delta z_j = z_j - z_{j-1}$ is the distance traveled in the $z$-direction, and $z_0$ is the initial $z$-coordinate. 

The values of $B_m$ and $\beta_m$ depend on the specific random scattering path $\gamma$, but we can statistically estimate their distributions. In a medium with scattering coefficient $\mu_s$ the distance between scattering events can be modeled as
\begin{align} \label{eq:Delta_L}
    \Delta L_j = -\ln(\xi_j) / \mu_s,
\end{align}
where $\xi_j$ is a uniform random value from the open interval $(0, 1]$. In general, $\Delta z_j$ is given by 
\begin{align}
    \Delta z_j &= \hat{\kappa}_{j, z} \Delta L_j,
\end{align}
where $\hat{\kappa}_{j, z}$ is the $z$ component of the normalized light wave vector for the $j^{\text{th}}$ straight line. Furthermore, we assume that $z_0$ is equally likely to lie anywhere on the ultrasound oscillation, i.e., $z_0 \in [0, 2\pi/k_{m, z}]$, which should be fairly accurate in a real case where the ultrasound focus propagates through the medium and where the drop-off in pressure is gradual.

For isotropic scattering 
\begin{align} \label{eq:Delta_z}
    \hat{\kappa}_{j, z} = 2\zeta_j - 1,
\end{align}
where $\zeta_j \in $ [0, 1]. However, for tissue, the Henyey-Greenstein phase function is more commonly used \cite{Prahl1989}, where the random scattering angles $\varphi$ and $\theta$ between successive scattering events are given by 
\begin{align}\label{eq:phi_and_theta}
    \varphi_j \in [0, 2\pi], \quad \cos(\theta_j) = \frac{1}{2g} \left( 1 + g^2 - \left( \frac{1 - g^2}{1 - g + 2g\zeta_j}\right)^2 \right), 
\end{align}
where $g$ is the anisotropy coefficient. Assuming a random initial direction of the photon, $\hat{\kappa}_{j, z}$ can be calculated through the recursive formula
\begin{align}
    \hat{\kappa}_{j, z} = -\sin(\theta_j) \cos(\varphi_j)\sqrt{1 - \hat{\kappa}_{j-1, z}^2} + \hat{\kappa}_{j-1, z}\cos(\theta_j), \quad \hat{\kappa}_{0, z} = 2\zeta_0 - 1. 
\end{align}

Note that estimating the distributions in this way can generate paths that extend outside the ultrasound volume, but the overall statistical results still yield good agreement as will be shown in Sec. \ref{sec:results}. 

Finally, $B_m$ and $\beta_m$ depend on the number of straight paths $J$, which thus must be determined. For a given photon path that travels a length $L$ within the ultrasound volume, the average scattering distance of $1/\mu_s$ can be used to estimate $J \approx L \mu_s$. Furthermore, if only the average photon path length $\langle L \rangle$ is known for a given ultrasound volume and scattering coefficient, we can use the probability density function of obtaining a specific $L$
\begin{align}\label{eq:P_L}
    \mathcal{P}(L) = \frac{1}{\langle L \rangle} e^{-L/\langle L \rangle},
\end{align}
to estimate the distribution of $J$ values according to 
\begin{align}\label{eq:P_J}
    P(J) \propto \frac{1}{\langle L \rangle} e^{-J/(\mu_s \langle L \rangle)}, \quad \sum_J P(J) = 1. 
\end{align}

\begin{figure}
\includegraphics[width=\columnwidth]{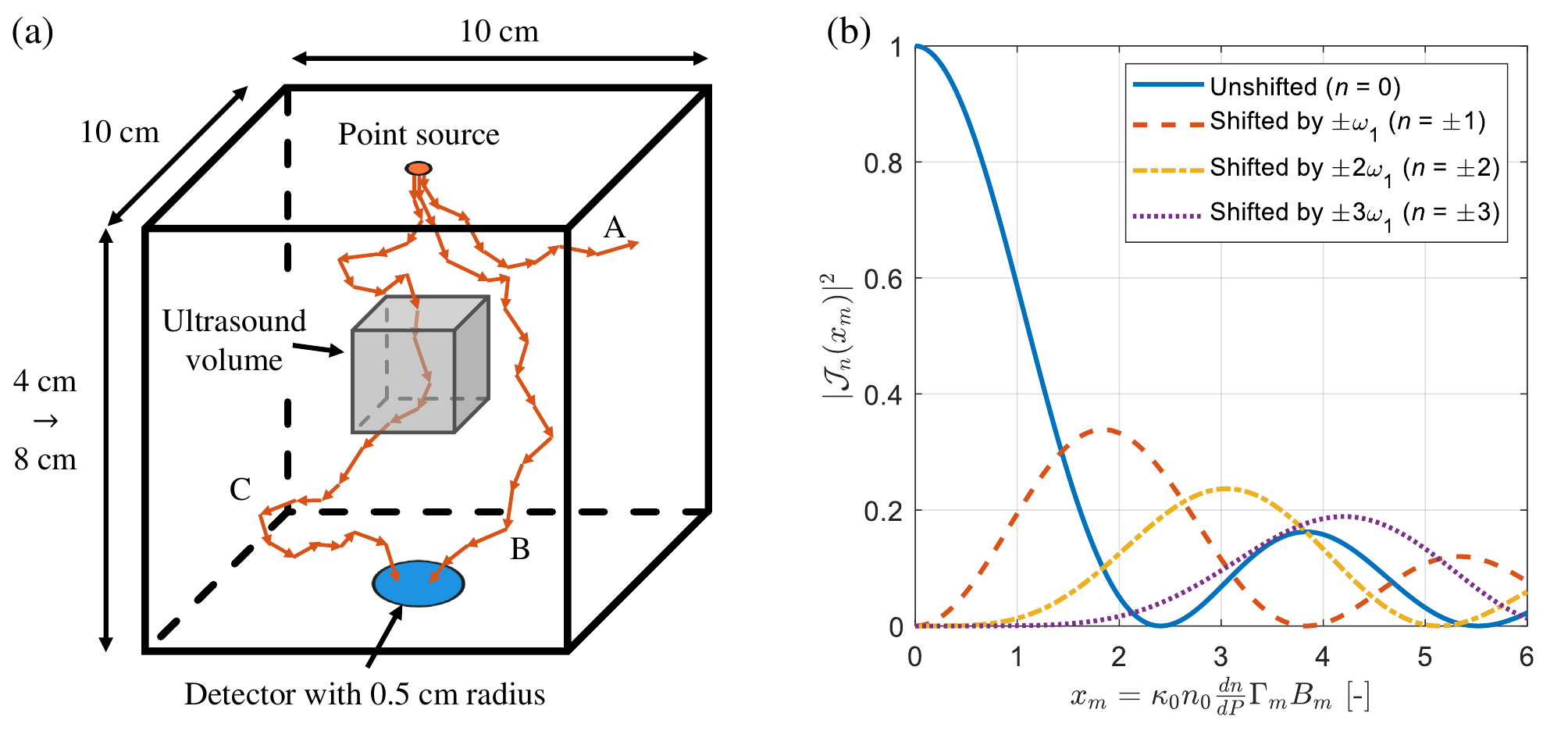}
\caption{\label{fig:Monte_Carlo}(a) Shows the Monte Carlo simulation volume, where all photons are initialized at a point source (red circle at the top). Each photon falls into one of three categories: A) the photon leaves the simulation volume without being detected, B) the photon is detected but does not enter the ultrasound volume, or C) the photon is detected and has been inside the ultrasound volume. New photons are initialized until $10,000$ photons fall into either category B or C. However, only photons in category C are analyzed when investigating the light-ultrasound interaction. (b) Shows the oscillatory behavior of the Bessel functions. Each photon path has a specific value of $x_m$, and the corresponding $|\mathcal{J}_n(x_m)|^2$ represents the probability of being frequency-shifted by $n\omega_1$.}
\end{figure}

\begin{figure*}
\includegraphics[width=\textwidth]{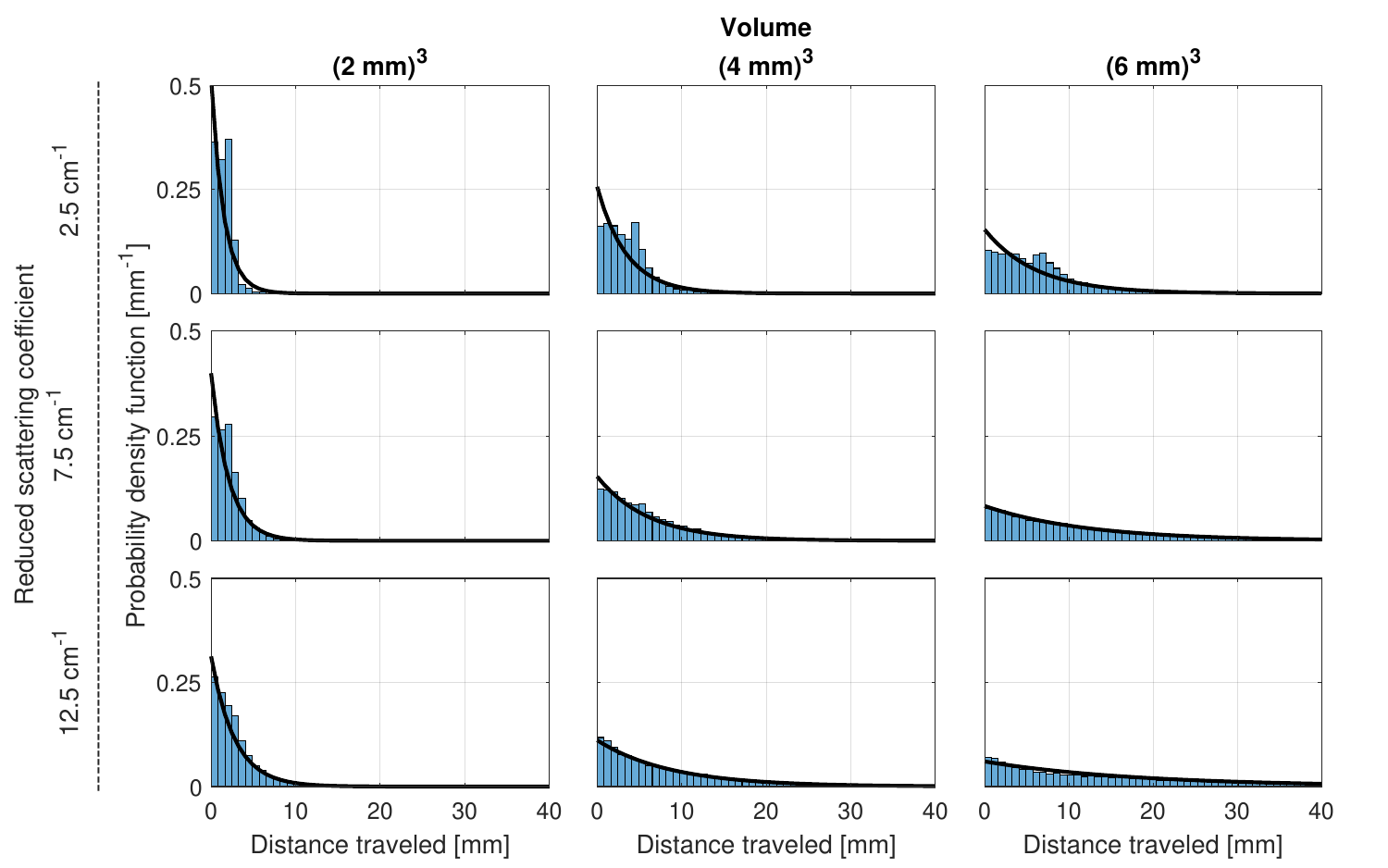}
\caption{\label{fig:D_distributions}Shows the probability density function for the distance traveled inside the ultrasound volume for different volumes and reduced scattering coefficients. The histogram only includes photons that have a non-zero distance within the ultrasound volume. The material has no absorption and the anisotropy coefficient is $g = 0.9$. The blue bars show the Monte Carlo results, whereas the black lines show the results from Eq. (\ref{eq:P_L}) using the average path length from the Monte Carlo simulations. }
\end{figure*}

\section{Analytic model validation using Monte Carlo simulations}\label{sec:results}
To validate the analytical model, we compare the results with Monte Carlo simulations where computational photons scatter according to Eqs. (\ref{eq:Delta_L}) and (\ref{eq:phi_and_theta}). We use the sequential Monte Carlo method, as presented in Chapter 3.3 of Ref. \cite{Hill2022}, and an open-source version of the code is discussed in Ref. \cite{Hill2021a}. In this method, computational photons scatter throughout the simulation volume according to Monte Carlo light transport until the photons exit the simulation boundary. Unlike many other Monte Carlo simulations, light absorption is not included at this stage. Instead, photons that exit through a predefined detection area have their paths saved. Absorption is introduced afterward by dividing the simulation volume into $V$ voxels, each with a distinct absorption coefficient $\mu_{a, v}$, and then for each saved photon path the total distance traveled within each voxel, $D_v$, is calculated, and the photon transmission is determined through the Bouguer-Beer-Lambert law: $T=e^{-\sum_v D_v \mu_{a, v}}$. The advantage of this method is that, as long as the scattering remains unchanged, the same computational photons can be reused to quickly calculate the transmission for different absorption settings. For further details, see Refs. \cite{Hill2022, Hill2021a}. In our current work, the sequential Monte Carlo method allows us to verify our analytical light-ultrasound interaction method without considering absorption, as it can be added later as explained above.

The simulation volume was $10\times10$ cm$^2$ wide and the thickness varied between $4$ and $8$ cm in steps of $1$ cm, see Fig. \ref{fig:Monte_Carlo}(a). The photons were initialized in the center of the top surface of the volume, traveling straight down, and scattered until they left the volume. If the photons exited at the detector, which we define as a $0.5$ cm radius circle placed straight underneath the source at the bottom of the volume, the photon paths were saved. For each simulation with a different thickness or scattering coefficient, we initialized new photons until $10,000$ photon paths were detected and saved. The thickness did not significantly alter the distribution of photons entering the ultrasound volume, which was placed in the center of the simulation volume. Therefore, in the following analysis, we show the results of all thicknesses combined to reduce the statistical error, i.e., $50,000$ saved photon paths are examined for each scattering coefficient. The detector radius is relatively arbitrary but was chosen to be large enough to collect sufficiently many photons within a reasonable time, without being too large as that would result in fewer photons entering the ultrasound volume. Since these simulations aim to validate the analytical light-ultrasound interaction method, only the distance the photons travel within the ultrasound volume is important, and the main results of this article are therefore unaffected by the size and placement of the detector.

In Fig. \ref{fig:D_distributions} we show the distributions of total path lengths traveled within the ultrasound volume for several different ultrasound volumes and scattering coefficients. In each case, the distribution only includes photons that have a non-zero total path length within the ultrasound volume, i.e., category C in Fig. \ref{fig:Monte_Carlo}(a). In all simulations, the material was homogeneous with an anisotropy coefficient of $g = 0.9$. As shown, the exponential distribution of Eq. (\ref{eq:P_L}) fits the data very well, except for the cases where the scattering coefficient or the volume are small, since then many photons can travel straight through the volume resulting in a peak in the Monte Carlo distribution around the size of the volume. Throughout this work, the analytical model will use the exponential distribution of path lengths according to Eq. (\ref{eq:P_L}). 

\begin{figure*}
\includegraphics[width=\textwidth]{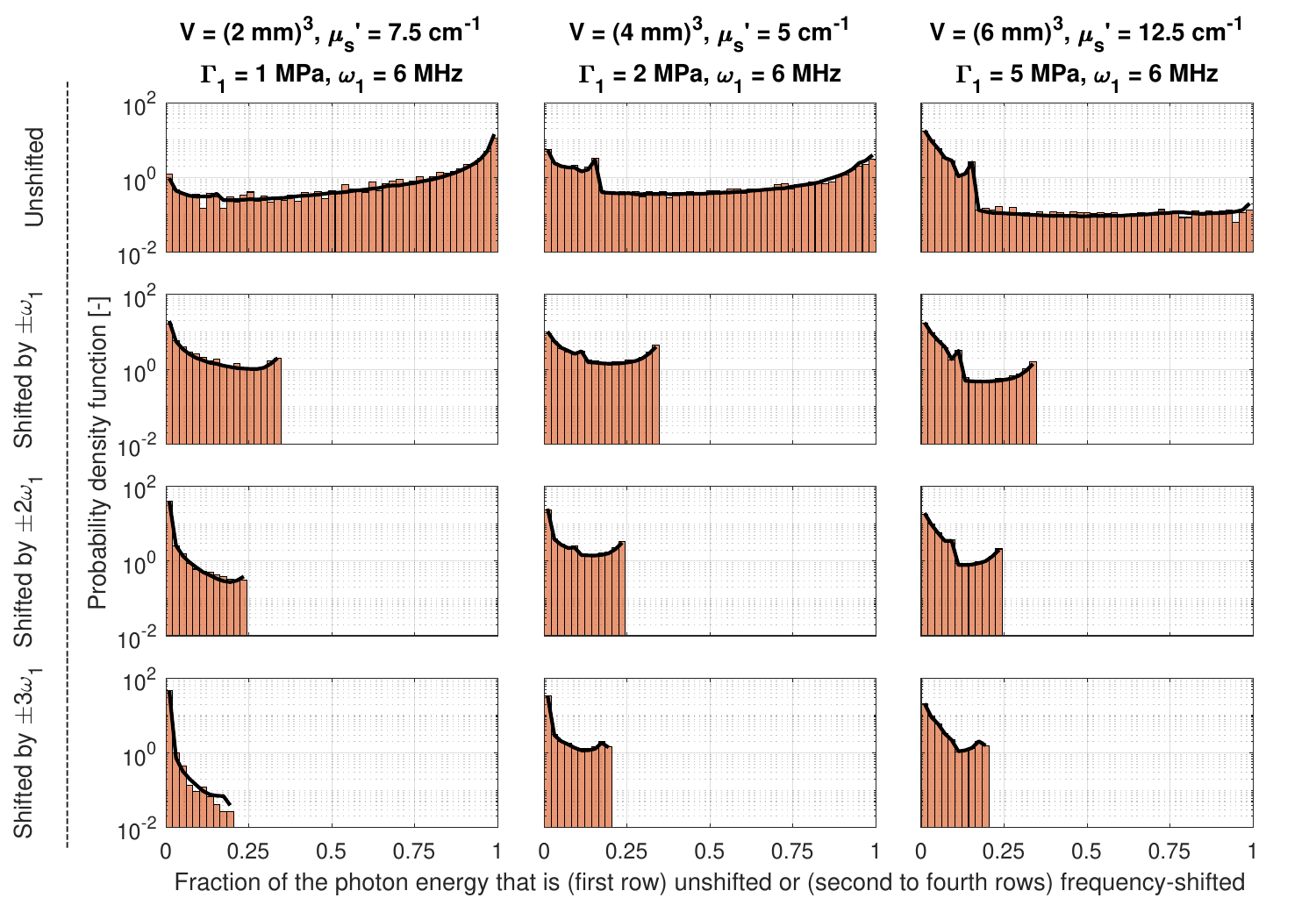}
\caption{\label{fig:A_histograms}Shows the probability density function that a certain fraction of the photon energy is (first row) not frequency-shifted, or (second to fourth rows) frequency-shifted by $1\rightarrow 3$ times $\pm\omega_1$, for different material and ultrasound properties (columns). Red bars show the results from Monte Carlo simulations, where the frequency content of each photon is calculated by numerically evaluating the equations in Sec. \ref{sec:theory}. The black lines are calculated using Eq. (\ref{eq:I_rel_weighted}).}
\end{figure*}

The light-ultrasound interaction was then simulated for all Monte Carlo photon paths following the procedure outlined in Refs. \cite{Hill2021, Hill2022}, which numerically evaluates Eqs. (\ref{eq:pressure})-(\ref{eq:electric_field}), then Fourier transforms the electric field, before integrating the results to obtain the probability that a photon becomes frequency-shifted. Figure \ref{fig:A_histograms} shows the results for a few different ultrasound and material properties. In all cases, the ultrasound pressure was described by Eq. (\ref{eq:pressure}), which filled a cubic volume, $V$, of varying size located at the center of the simulation volume. The wavelength of light was $689$ nm, since there exists a promising rare-earth filter at that wavelength which can be used for ultrasound optical tomography \cite{Bengtsson2022a}. The background index of refraction was $n_0 = 1.3$, and we used the piezo-optic coefficient of water, $\frac{dn}{dP} \approx 1.466\times10^{-10}$ Pa$^{-1}$ \cite{Riley1967}, and the ultrasound speed in 20$^{\circ}$ C water, $1480$ m/s \cite{VandeSompel2012}. 

Figure \ref{fig:A_histograms} also validates the procedure outlined in Secs. \ref{sec:frequency_content} and \ref{sec:path_integral}, where the relative light intensities at the different frequencies $n\omega_1$ are given by Eq. (\ref{eq:I_rel}), weighted by the probability, $P(J)$, that a photon scatters $J$ times as given in Eq. (\ref{eq:P_J}). Since only one ultrasound frequency is used, the relative light intensities can be simplified to 
\begin{align}\label{eq:I_rel_weighted}
    I_\text{rel}(n) = \sum_J P(J) \left| \mathcal{J}_n(x_m) \right|^2,
\end{align}
where $x_m$ depends on $B_m$ which depends on $J$ and the random variables $\xi_j$, $\zeta_j$, and $\varphi_j$, hence $x_m$ is different for each photon path. Since we are interested in the overall statistical results, we generate $10,000$ $B_m$ values for each $J$. Furthermore, we cap the summation at $J = \lceil 7 \mu_s \langle L \rangle \rceil$. 

Since Bessel functions of negative order relate to the positive order via $\mathcal{J}_{-n}(x_m) = (-1)^n \mathcal{J}_n(x_m)$, we can see that positive and negative frequency shifts occur with the same probability if only one ultrasound frequency is present. However, this is not true in the general case as discussed in Sec. \ref{sec:multiple_US_freq}. 

The distinct sharp maximums in the probability density functions of Fig. \ref{fig:A_histograms} are due to the oscillatory behavior of the Bessel functions as shown in Fig. \ref{fig:Monte_Carlo}(b). For example, the highest fraction of the photon energy that can be frequency-shifted by $\pm\omega_1$ as shown in the second row of Fig. \ref{fig:A_histograms} is given by the maximum value of $|\mathcal{J}_1(x_m)|^2$, which is around $0.34$ as shown in the dashed red curve in Fig. \ref{fig:Monte_Carlo}(b).

\begin{figure*}
\includegraphics[width=\textwidth]{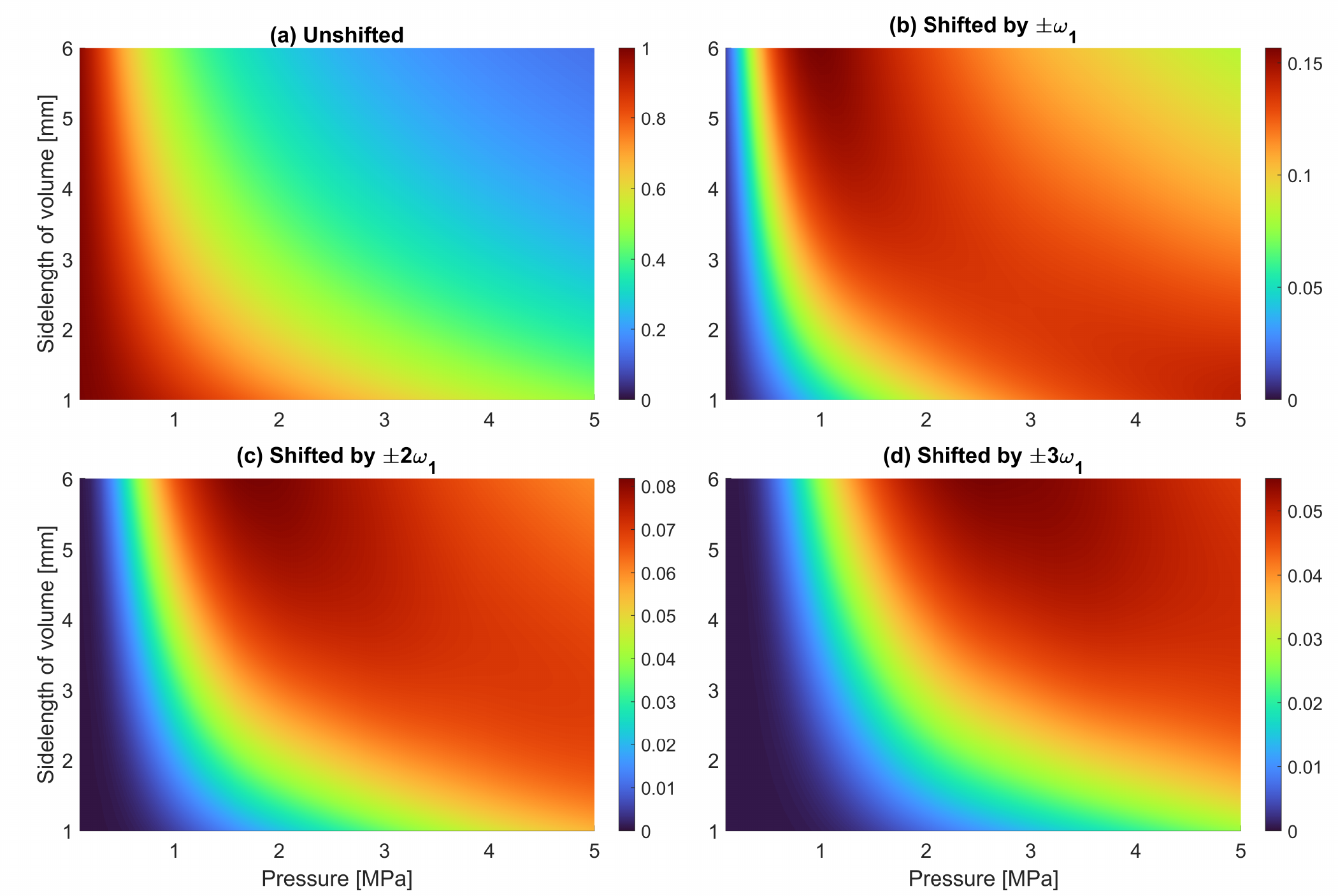}
\caption{\label{fig:Pressure_volume}Shows the average probability of being frequency-shifted as a function of the ultrasound pressure and volume. The ultrasound frequency is $\omega_1 = 6$ MHz and the reduced scattering coefficient is $\mu_s' = 7.5$ cm$^{-1}$. For each ultrasound volume, the average photon path lengths, $\langle L \rangle$, used in Eq. (\ref{eq:P_J}) are estimated based on Monte Carlo simulations. }
\end{figure*}

Due to this oscillatory behavior, increasing the ultrasound volume or pressure does not always lead to larger probabilities of being frequency-shifted into a specific harmonic of the ultrasound frequency, as can be seen in Fig. \ref{fig:Pressure_volume}. Instead, depending on the distribution of $x_m$ values, there exists an optimal ultrasound pressure for each ultrasound volume that maximizes the average probability that a photon is frequency-shifted by a specific amount, given that the photon has been inside the ultrasound volume. However, while this probability may decrease as the ultrasound volume increases, the larger volume enhances the likelihood of a photon entering the ultrasound volume in the first place, i.e., in Fig. \ref{fig:Monte_Carlo}(a), more photons are classified into category C. Consequently, when considering the total number of detected frequency-shifted photons, increasing the ultrasound volume is generally advantageous, especially when the fraction of photons entering the ultrasound volume is initially small.

\subsection{Multiple ultrasound frequency components}\label{sec:multiple_US_freq}
In the general case with multiple ultrasound frequency components, the probabilities of achieving negative or positive frequency shifts can differ due to the interference effects in Eq. (\ref{eq:I_rel}), which contrasts with the one ultrasound frequency case where the probabilities are equal. These interference effects are analog to those occurring in an electro-optical modulator (EOM), except then the interaction occurs with one strength, whereas in our case the interaction occurs with a distribution of strengths $x_m$ that depend on the photon path integral discussed in Sec. \ref{sec:path_integral}. 

\begin{figure*}
\includegraphics[width=\textwidth]{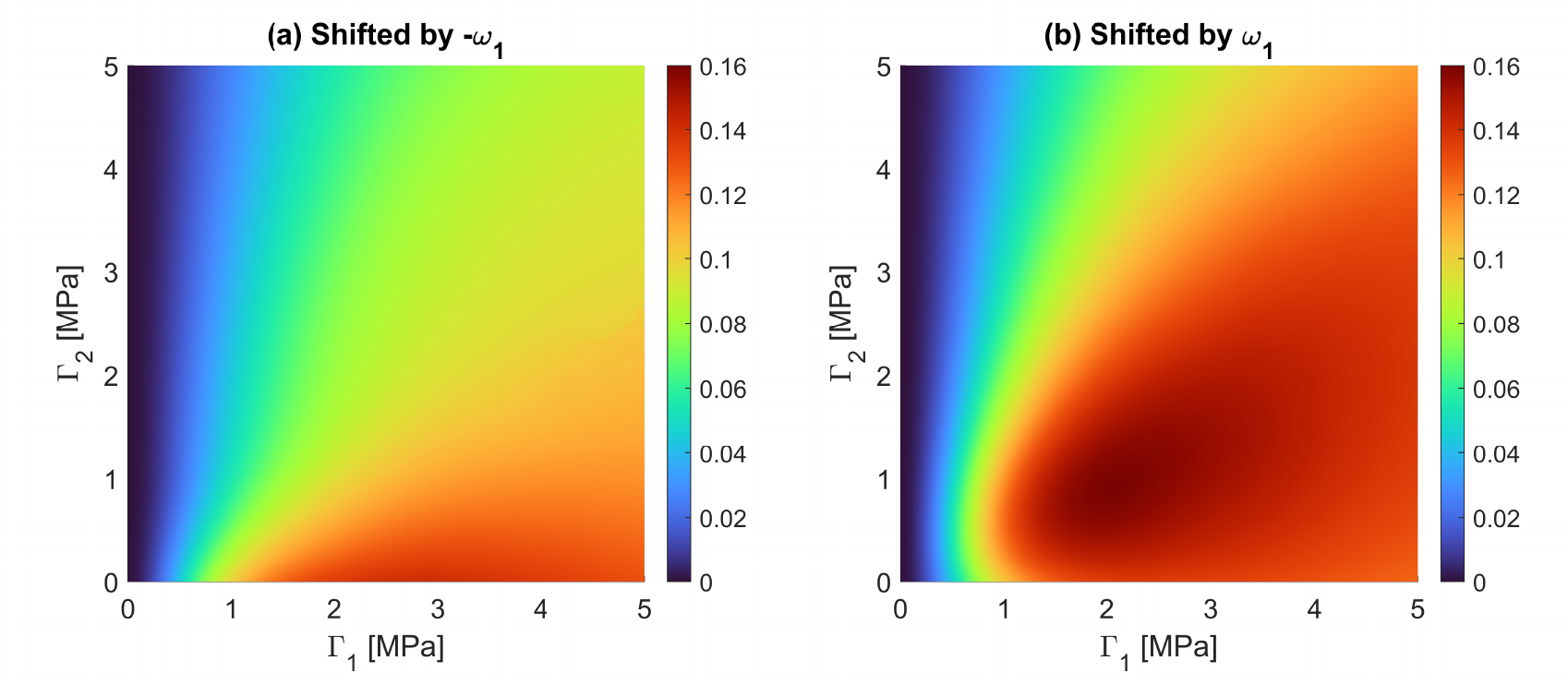}
\caption{\label{fig:Pressure_optimum}Shows the average probability of being frequency-shifted by $\pm\omega_1$ as a function of the first and second order harmonic pressures, $\Gamma_1$ and $\Gamma_2$. For each configuration of pressures, the average probability of being shifted by $+\omega_1$ is maximized by optimizing the phase of the second ultrasound harmonic. The sidelength of the ultrasound volume is $3$ mm, and as in Fig. \ref{fig:Pressure_volume} the fundamental ultrasound frequency is $\omega_1 = 6$ MHz and the reduced scattering coefficient is $\mu_s' = 7.5$ cm$^{-1}$.}
\end{figure*}

Figure \ref{fig:Pressure_optimum} shows the average probability of being frequency-shifted by $\pm\omega_1$ as a function of the pressures of the first and second harmonics, where the phase of the second harmonic has been optimized to maximize the $+\omega_1$ frequency-shift. For ultrasound waves that are harmonics of a fundamental frequency, a translation in time or space alters the phases $\alpha_m$ by $m\alpha_1$, where $\alpha_1$ is the phase change of the fundamental wave, and such phase changes generate the same interference effects. Thus, if only the fundamental and the second harmonic are present, it is enough to optimize the phase of the second harmonic as we have done. As can be seen from this simple example, it is possible to achieve a higher average probability when the second harmonic is present compared to when it is not. 

\begin{figure*}
\includegraphics[width=\textwidth]{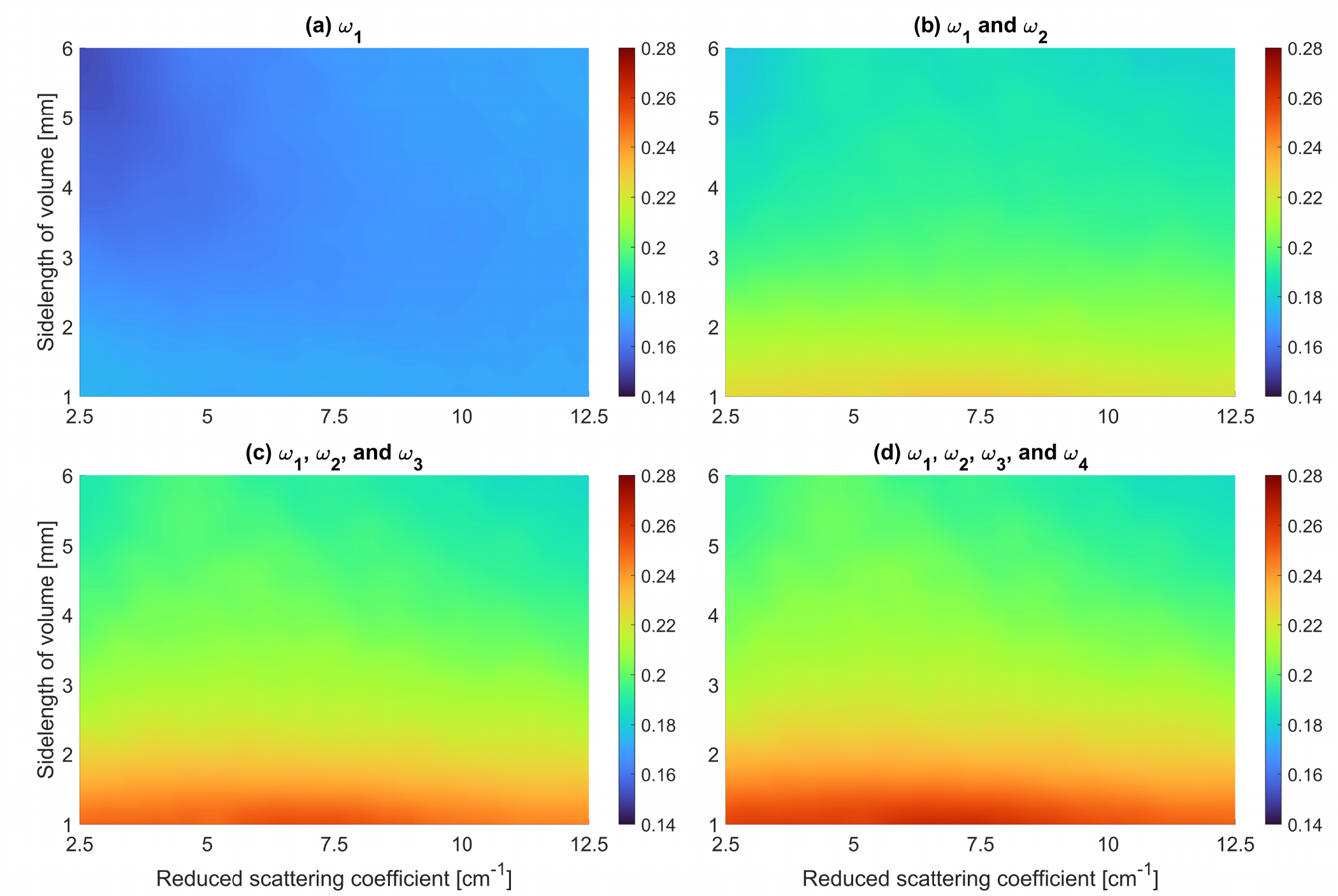}
\caption{\label{fig:Optimized_tagging}Shows the average probability of being frequency-shifted by $+\omega_1$ as a function of the reduced scattering coefficient and ultrasound volume. In (a) only the fundamental ultrasound frequency $\omega_1$ is present, whereas in (b-d) additional harmonics are added. The pressure and phase of the harmonics are related via $\Gamma_m = \Gamma_1/m$ and $\alpha_m = (m-1)\pi$, and $\Gamma_1 \in [0, 5]$ MPa is optimized for each setting to maximize the probability. The fundamental ultrasound frequency is $\omega_1 = 1$ MHz, and to speed up the calculations, we only consider $n_m \in [-3, 3]$. For each setting, the average photon path lengths, $\langle L \rangle$, used in Eq. (\ref{eq:P_J}) are estimated based on Monte Carlo simulations. }
\end{figure*}

If only one interaction strength $x_m$ had been present, one can optimize the probability of being frequency-shifted by $+\omega_1$ by relating the properties of the ultrasound harmonics via $\Gamma_m = \Gamma_1/m$ and $\alpha_m = (m-1)\pi$, and only optimize $\Gamma_1$. Despite having a distribution of interaction strengths, this approach still increases the probability in our case, as shown in Fig. \ref{fig:Optimized_tagging}. However, due to the large distribution of $x_m$, the gains are relatively modest, below a factor of $2$, at least for the cases shown in Figs. \ref{fig:Pressure_optimum} and \ref{fig:Optimized_tagging}. One should also consider the technical challenge to achieve such precise control of the ultrasound properties deep inside tissue due to the non-linear ultrasound propagation effects \cite{Duck2002} and the frequency-dependent attenuation of around $0.75$ dB/cm/MHz for soft tissue \cite{Shankar2011}. However, performing an optimization that accounts for this is beyond the scope of this article. 

\section{Conclusion}\label{sec:conc}
We have modeled how light is frequency-shifted when interacting with ultrasound within scattering media, resulting in simple analytical expressions involving summations of Bessel functions. The model is validated through Monte Carlo simulations and allows for the use of multiple high-pressure ultrasound frequency components, opening up the possibility of optimizing the pressures, frequencies, and phases of the ultrasound waves to maximize the probability that light is frequency-shifted. 

Even if such optimization only results in modest gains at this stage, the analytical framework presented here also has an advantage in general as it allows for much faster estimations of the light-ultrasound interaction compared to the numerical methods presented in Ref. \cite{Hill2021}, and non-linear ultrasound propagation effects in tissue demands the incorporation of several ultrasound frequency components, which previous analytical models were missing. 

Future research could expand on this model by incorporating the effects of scatterer movement due to the ultrasound waves, model time-varying or spatially-varying pressure fields, or investigate if more advanced ultrasound setups can result in larger probabilities of light being frequency-shifted.

\begin{backmatter}

\bmsection{Funding}
This work was partially supported by the Wallenberg Centre for Quantum Technology (WACQT) funded by Knut and Alice Wallenberg Foundation (KAW).

\bmsection{Acknowledgments}
We thank Stefan Kr\"{o}ll and Lars Rippe for their input and discussions. 

\bmsection{Disclosures}
AK: Deep Light Vision AB (E,P)

\bmsection{Data Availability Statement}
The data underlying the figures in this work were generated using both the analytical framework presented here and numerical Monte Carlo simulations based on Ref. \cite{Hill2022}, with an open-source version of the code discussed in Ref. \cite{Hill2021a}. As the data are derived from computational models and random input values, rather than experimental measurements, they are not publicly available.
\end{backmatter}

\bibliography{Ref_lib}

\end{document}